\documentclass[sigconf]{acmart}

\usepackage{todonotes}
\usepackage{balance}
\newcommand{\nop}[1]{}

\newcommand{\etal}{\emph{et~al. }}
\newcommand{\eg}{\emph{e.g., }}
\newcommand{\ie}{\emph{i.e., }}

\AtBeginDocument{%
  \providecommand\BibTeX{{%
    \normalfont B\kern-0.5em{\scshape i\kern-0.25em b}\kern-0.8em\TeX}}}

\setcopyright{acmcopyright}
\copyrightyear{2024}
\acmYear{2024}
\setcopyright{rightsretained}
\acmConference[SIGCSE 2024]{Proceedings of the 55th ACM Technical
Symposium on Computer Science Education V. 1}{March 20--23, 2024}{Portland, OR, USA}
\acmBooktitle{Proceedings of the 55th ACM Technical Symposium on Computer
Science Education V. 1 (SIGCSE 2024), March 20--23, 2024, Portland, OR, USA}
\acmDOI{10.1145/3626252.3630821}
\acmISBN{979-8-4007-0423-9/24/03}

\makeatletter
\gdef\@copyrightpermission{
  \begin{minipage}{0.3\columnwidth}
 \href{https://creativecommons.org/licenses/by-nc-nd/4.0/}{\includegraphics[width=0.90\textwidth]{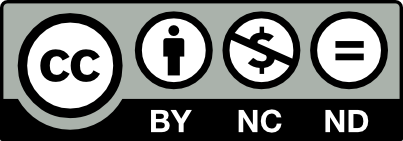}}  \end{minipage}\hfill
  \begin{minipage}{0.7\columnwidth}
   \href{https://creativecommons.org/licenses/by-nc-nd/4.0/}{This work is licensed under a Creative Commons Attribution-NonCommercial-NoDerivs International 4.0 License.}
  \end{minipage}
  \vspace{5pt}
}
\makeatother
\begin{document}

\title{Cybersecurity as a Crosscutting Concept Across an Undergrad Computer Science Curriculum: An Experience Report}

\author{Azqa Nadeem}
\affiliation{%
  \institution{Delft University of Technology \& University of Twente}
  \country{Enschede, Netherlands}
}
\email{a.nadeem@utwente.nl}

\begin{abstract}
Although many Computer Science (CS) programs offer cybersecurity courses, they are typically optional and placed at the periphery of the program. 
We advocate to integrate \textit{cybersecurity as a crosscutting concept} in CS curricula, which is also consistent with latest cybersecurity curricular guidelines, \eg CSEC2017.
We describe our experience of implementing this crosscutting intervention across three undergraduate core CS courses at a leading technical university in Europe between 2018 and 2023, collectively educating over 2200 students. 
The security education was incorporated within CS courses using a partnership between the responsible course instructor and a security expert, \ie the security expert (after consultation with course instructors) developed and taught lectures covering multiple CSEC2017 knowledge areas. 
This created a complex dynamic between three stakeholders: the course instructor, the security expert, and the students. 
We reflect on our intervention from the perspective of the three stakeholders -- we conducted a post-course survey to collect student perceptions, and semi-supervised interviews with responsible course instructors and the security expert to gauge their experience. 
We found that while the students were extremely enthusiastic about the security content and retained its impact several years later, the misaligned incentives for the instructors and the security expert made it difficult to sustain this intervention without organizational support. 
By identifying limitations in our intervention, we suggest ideas for sustaining it.

\end{abstract}




\maketitle
\section{Introduction}

In an increasingly interconnected world, where technology permeates every aspect of human lives, the demand for a secure cyberspace has become imperative. 
From 2013 to 2021, security-related jobs have increased by 350\% \cite{Morgan2023}. 
Meanwhile, there is a global shortage of cybersecurity workforce \cite{international2018cybersecurity} -- there are approximately 3.5M unfilled security vacancies in 2023 \cite{Morgan2023}. Hence, it is crucial to provide adequate cybersecurity training to the next generation in order to meet the needs of a cyber-ready workforce.

Although a number of dedicated cybersecurity degree programs have been developed for undergraduate \cite{blaine2021creating,lo2020information,tang2020curricular} and graduate levels \cite{cabaj2018cybersecurity,asghar2020case}, only a small fraction of the population follows such programs. Many Computer Science (CS) degrees fail to include cybersecurity as a major concern. As a result, security courses are typically offered as electives, which means that CS students can graduate without ever taking a security class \cite{cheung2011challenge,almansoori2021textbook}.

We propose to include \textit{cybersecurity as a crosscutting concept} across undergraduate CS curricula. This is based on the intuition that cybersecurity is naturally a crosscutting topic, just like quality assurance \cite{blaine2021creating,siraj2015integrating}. 
Thus, instead of developing a dedicated security course, the idea here is to infuse cybersecurity topics within existing course materials.
This approach is also supported by the Cybersecurity Curricula 2017 (CSEC2017) \cite{force2017cyber} that identifies several security aspects as crosscutting concerns.  
There are four benefits to this approach: (i) Security instructors do not have to compete for a place in the curriculum. (ii) Introducing security in core CS courses ensures that all CS graduates have basic security knowledge. (iii) Introducing security early-on and revisiting it often helps develop \textit{`the security mindset'} \cite{siraj2021there}. (iv) Teaching CS topics together with their security implications ensures that security is no longer considered an after-thought. 

In this report, we describe our experience of implementing cybersecurity as a crosscutting concept across \textit{three core courses} within an undergraduate CS curriculum over a period of \textit{five years}, providing education to \textit{over 2200 students} at a leading technical university in Europe. 
%
Prior research \cite{blair2020infusing} identifies the major challenges of integrating cybersecurity into CS programs as uncertainty about \textit{what to teach}\footnote{Teaching material that infuses security with CS topics is not readily available -- textbooks rarely discuss the security implications associated to CS topics \cite{almansoori2021textbook}.} and \textit{who can teach it}. 
We address the former challenge by selecting topics covering a wide range of Knowledge Areas (KAs) proposed in the CSES2017. We address the latter challenge by having cybersecurity experts teach the security content, \ie lecturers with a security background. 
In order to include security as an explicit learning objective in each course, the course instructor and the security expert work together to select a suitable security topic, while the security expert develops, delivers, and maintains the learning materials. This creates a complex dynamic between three stakeholders: the responsible course instructor, the security expert, and the students. 

Our crosscutting intervention is organized as follows: Once a security topic is finalized for a course, the security expert develops a lecture, an accompanying hands-on assignment, and exam questions.
Because students are required to understand the basic topical concepts before understanding where security plays a role, the security lectures are placed towards the end of the course. In one or more 45-minute lecture slots, the security expert introduces the security topic, drawing in the relevant concepts they have studied throughout the course. The students also have access to a hands-on activity during the lecture to aid their learning. Finally, student learning is tested via several scenario-based exam questions. 
This set up has been used to provide security education to first year undergraduate students from 2018 to 2023.

We evaluate the crosscutting intervention from the perspective of the three stakeholders, \ie student perception, course instructor reflection, and security expert experience. To this end, we conduct a post-course survey with year-1 students (to evaluate the fit and suitability of the security lecture), and with year-2/3 students (to gauge interest retention in security). We also conduct semi-supervised interviews with the course instructors and the security expert to get an insight into their experience with the proposed intervention.

Our results suggest that there exists a misalignment of incentives that makes it extremely difficult to sustain such a three-stakeholder-based crosscutting intervention. While the students are extremely enthusiastic about the security content, there is a power struggle between the course instructors and the security expert with respect to adding security content \textit{vs.} retaining existing course content, which is further exacerbated by the participation of several courses. This makes it difficult to sustain such an effort without organizational oversight, which was also observed by Petel \etal \cite{petelka2022principles}. 
We provide recommendations for making this intervention more sustainable.

\section{Related work and background}

Goupil \etal have identified that the shortage of a cyber-ready workforce is predominantly caused by the misalignment of the skill-demand of the jobs and the inadequate qualifications possessed by the applicants \cite{goupil2022towards}. They found that the discrepancy between supply (candidate skills) and demand (job vacancies) is a result of insufficient topic coverage in computer science (CS) curricula.

If \textit{``good security practices are simply good computing practices''} \cite{blair2020infusing}, then we should incorporate cybersecurity within existing CS curricula to nurture more effective CS graduates who also have the security mindset. 
There are several efforts from governing bodies to make security a crosscutting concept. For instance, the 2013 ACM/IEEE-CS curricula report \cite{joint2013computer} designated \texttt{Information Assurance and Security (IAS)} as a new crosscutting Knowledge Area (KA). Later, the Joint Task Force on Cybersecurity Education proposed the Cybersecurity Curricula (CSEC2017) \cite{force2017cyber} that explicitly makes several security aspects crosscutting concepts, \eg  \texttt{Adversarial thinking}, \texttt{Confidentiality}, and \texttt{Risk}. In our intervention, we go one step further and convert several KAs as crosscutting concepts as well, \ie \texttt{Data security}, \texttt{Software security}, \texttt{Component security}, and \texttt{Connection security}.

A limited number of studies have investigated the impact of cybersecurity as a crosscutting concept. 
For instance, Siraj \etal \cite{siraj2015integrating} present a toolkit that aims to develop faculty expertise in cybersecurity so that they can include cybersecurity education throughout an undergraduate CS curriculum, while Blair \etal \cite{blair2020infusing} propose curricular guidelines for infusing secure computing in an existing undergraduate CS curriculum.
In contrast, we describe our practical experience of implementing security as a crosscutting concept across three courses over a period of five years. 
Recognizing that it will take some time before existing teaching staff is trained to become well-versed in cybersecurity \cite{mourning2022reflections,siraj2015integrating}, we include the security expert as an added stakeholder within CS education. The benefit of including security experts is that they understand the changing threat landscape, and are aware of critical security topics.

Cybersecurity is a highly practical domain that requires nuanced teaching techniques. Existing work has investigated the use of educational theories and practical lab-work to enhance the effectiveness of security education. For instance, prior work investigates challenge-based learning \cite{cheung2011challenge}, spiral theory \cite{basu2020integration}, collective concept maps \cite{perouli2021assessing}, serious games \cite{kletenik2020cyber}, and POGIL activities \cite{mezei2020introducing} to teach security. Moore \etal \cite{moore2022cybersecurity} even use card magic to teach randomness for cryptography.  
Other works implement practical hands-on labs to teach various security topics, \eg web security \cite{shar2022xss,weinfurter2021raising,deng2021neocyberkg}, mobile/wireless security \cite{oconnor2021teaching}, malware detection \cite{lo2020hands}, fraud detection \cite{shahriar2020case}, and security testing \cite{james2020hands}. Capture The Flag (CTF) style exercises are a common way to teach cybersecurity \cite{ellis2021experience,cole2022impact}. 
We have taken inspiration from many of these works to develop the practical hands-on component of our security lectures.

\begin{table*}[t]
\caption{CSLS overview: The security content, the CSEC2017 KAs, and their placement within year-1 courses.}
\label{tab:csls-overview}
\resizebox{2\columnwidth}{!}{%
\begin{tabular}{ccccc}
\toprule
\textbf{Course} & \textbf{Quarter} & \textbf{Security Topic} & \textbf{Knowledge Area} & \textbf{Lecture Content/Learning Objectives} \\ \hline
\begin{tabular}[c]{@{}c@{}}Web and Database \\ Management (WDT)\end{tabular} & Q2 & Web security & \begin{tabular}[c]{@{}c@{}}Connection Security,\\ Software Security\end{tabular} & \begin{tabular}[c]{@{}c@{}}Recent security incidents; OWASP Top-10 and their mitigation; Demo using \\OWASP Juice Shop \cite{owasp2023} \end{tabular} \\ \hline
\begin{tabular}[c]{@{}c@{}}Information and Data \\ Management (IDM)\end{tabular} & Q3 & Data security & \begin{tabular}[c]{@{}c@{}}Data Security,\\ Software Security\end{tabular} & \begin{tabular}[c]{@{}c@{}}Recent security incidents; Threats to information security; Securing data at rest \\ and in motion; SQL injection -- anatomy and types; Mitigation -- data sanitization, \\ escaping, and prepared statements; Demo using a custom web app\end{tabular} \\ \hline
\begin{tabular}[c]{@{}c@{}}Software Quality \\ and Testing (SQT)\end{tabular} & Q4 & Security testing & \begin{tabular}[c]{@{}c@{}}Software Security,\\ Component Security\end{tabular} & \begin{tabular}[c]{@{}c@{}}Recent security incidents; Java vulnerabilities; Secure software development life \\ cycle; SAST vs. DAST; Static testing -- risk analysis, syntax analysis, structural \\ analysis; Dynamic testing --  tainting, fuzzing, dynamic validation, penetration testing\end{tabular} \\ \bottomrule
\end{tabular}}
\end{table*}

\section{Crosscutting Security Curriculum}
In this section, we provide an overview of the curriculum where we embed cybersecurity, and explain our intervention design. 

\paragraph{BSc curriculum overview.} 
Delft University of Technology is a leading technical university in the Netherlands.
It offers Computer Science and Engineering (CSE) as a three-year undergraduate degree, which contains 180 points in the European Credit Transfer and Accumulation System (ECTS).
Year-1 and three quarters of year-2 consist of compulsory courses. The rest of the program consists of elective courses, a minor, and a research project (\ie BSc thesis). 
Over the first two years, the students must follow 21 courses (18 compulsory, and 3 out of 9 variant courses).
Each course runs for a quarter (10 weeks), and is worth 5 ECTS. The courses are evaluated yearly via student surveys.

At the start of the intervention in 2018, there was zero explicit security content within the BSc CSE program, which is why we decided to incorporate security in compulsory CS courses. In 2019, a year-3 elective course on cybersecurity was offered, but the majority of the graduating students did not follow it, and effectively graduated without any security knowledge.

\paragraph{Intervention design.} We introduce security education within year-1 BSc CSE courses in the form of CyberSecurity Lecture Series (CSLS).
What makes our implementation of CSLS unique is that \textit{security experts} teach the security content in different courses. This enables them to amplify their reach to more students as opposed to being responsible for one standalone security course. 
The author of this report served as the security expert. She was hired as a cybersecurity PhD researcher with teaching responsibilities.

The idea was to include security as a Learning Objective (LO) in each course. The responsible instructor and the security expert worked together to select a suitable security topic for a course. The security expert developed, delivered, and maintained the learning materials, which were appropriately embedded within the corresponding course content. 
For each course, the learning materials included an lecture covering one to four 45-minute lecture slots, an accompanying hands-on assignment, and a number of exam questions for the midterm/final exam. 
The lectures were taught by the security expert towards the end of the course. This enabled the students to have an overview of the computing topic before understanding the role of security. We evaluated student learning with scenario-based open-ended and multiple-choice exam questions.

Cybersecurity is a field with a constant arms-race between attackers and defenders, which causes the threat landscape to evolve rapidly. 
Our CSLS lectures start with examples of recent attacks for the students to grasp the urgency and impact of the topic. In order to inculcate a security mindset among the students, the lectures also have an offensive part (to show how the attacks can be done), and a defensive part (to show how they can be mitigated). They also discuss responsible disclosure of vulnerabilities. 
To make the lectures more impactful and engaging, they have an in-class hands-on component so the students can practice the attacks and their mitigation.  
As such, the lectures include the following \textit{crosscutting concepts} from CSEC2017: \texttt{Adversarial thinking}, \texttt{Confidentiality}, \texttt{Integrity}, and \texttt{Availability}. 

The long-term intent is to make CSLS worth 5 ECTS, which when combined, is equivalent to a full undergraduate course. At the time of reporting, we have successfully integrated cybersecurity within three year-1 core courses, see Table \ref{tab:csls-overview} for an overview. We have included \textit{Web security} in CSE1500 (Web and Database Technologies), \textit{Data security} in CSE1505 (Information and Data Management), and \textit{Security testing} in CSE1110 (Software Quality and Testing). These topics were also identified by \cite{goupil2022towards} as having a higher industry demand but a lower curricular coverage.  
The courses are placed within year-1 Q2, Q3, and Q4. We selected these courses (to start with) since they are compulsory, and  the students have already had a basic introduction to CS in Q1. Other topics, \eg AI security and email security will be added later. The lecture contents can be accessed via \textit{\url{https://azqanadeem.github.io/teaching/}}.

CSLS has been offered to over 2200 undergraduate students over five years. Table \ref{tab:student-numbers} shows a breakdown of the number of students who were enrolled from 2018 to 2023 (who submitted the final exams of the three corresponding courses). We use this as a proxy for student attendance since we do not explicitly log attendance, and the lectures are often recorded for offline viewing.

\begin{table}[t]
\centering
\caption{Student enrollment figures who received security education as a crosscutting concept over five years.}
\label{tab:student-numbers}
\begin{tabular}{cccc}
\toprule
\textbf{Started in $\downarrow$} & \textbf{WDT (Q2)} & \textbf{IDM (Q3)} & \textbf{SQT (Q4)} \\ \midrule
\textbf{2018--2019} & 686 & 582 & 600 \\
\textbf{2019--2020} & 406 & 393 & 383 \\
\textbf{2020--2021} & 409 & 408 & 388 \\
\textbf{2021--2022} & 426 & 416 & 413 \\ 
\textbf{2022--2023} & 445 & 415 & 417 \\ \midrule
\textbf{Total students} & 2372 & 2214 & 2201 \\ \bottomrule
\end{tabular}
\end{table}

\subsection{Security Testing in SQT (CSE1110)} 
CSE1110 is a Q4 course that covers software testing techniques in order to prepare students for building high quality software systems \cite{aniche-pragmatic-testing-education}. The students learn to program primarily in Java.
We cover the \texttt{Software Security} and \texttt{Component Security} KAs, and include a lecture on \textit{Security Testing} in week 8. The learning objective (LO) is to consider testing from an adversarial perspective, and to compare Static Application Security Testing (SAST) and Dynamic Application Security Testing (DAST) techniques. The lecture also dissects popular Java exploits to understand how they work. The lecture is linked with the rest of the course via its examples in Java, and a discussion of concepts, \eg abstract syntax trees, data flow analysis, and code coverage. 
The security lecture was allocated 180 minutes when it was first introduced in 2018. It was allocated only 45 minutes in 2023. We discuss the reasons in Section \ref{sec:instructor-reflection}.%

\begin{figure}[t]
    \centering
    \includegraphics[width=\linewidth]{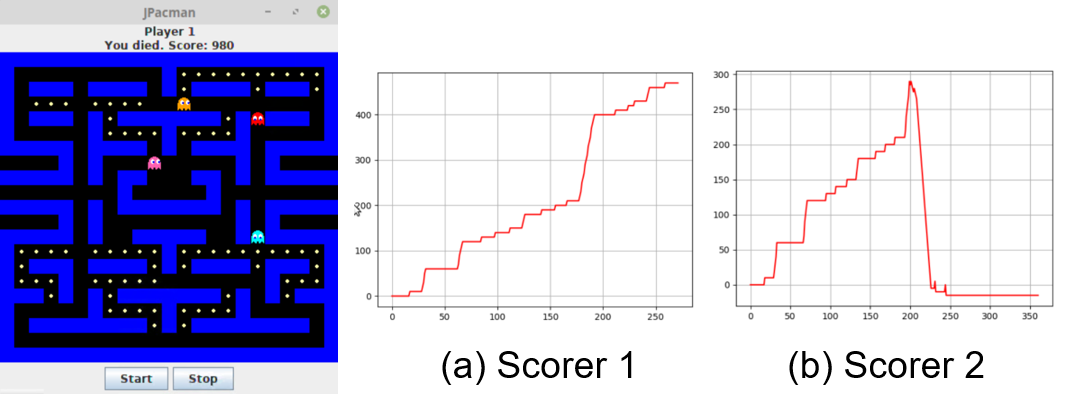}
    \caption{A buggy Pacman scoring module for SQT.}
    \label{fig:sqt-assignment}
\end{figure}

The hands-on assignment is part of a larger assignment where the students are tasked to develop Pacman in Java. We provide them with two scoring modules for the game, one of which is buggy. We also provide a fuzzing agent for automating game-play and creating logs. The students work in pairs to pinpoint the trigger of the misbehaving scoring module using SAST \textit{vs.} DAST, see a snapshot of the game in Figure \ref{fig:sqt-assignment}. 

\subsection{Data security in IDM (CSE1505)}
CSE1505 is a Q3 course that discusses data management and the implementation of relational databases, such as MySQL. 
We cover the \texttt{Data Security} and \texttt{Software Security} KAs, and include a lecture on \textit{Data Security} during week 9. The LO is to discuss the principles of information security, and techniques for keeping data confidential at rest and in motion. The lecture discusses the SQL injection (SQLi) attack in depth -- how it concretely works, its mitigation, and the responsible stakeholders. In a 90-minute lecture, the students learn to implement SQLi as a means to steal sensitive data stored in databases, and to mitigate this vulnerability.

For the hands-on component, we set up a vulnerable web shop hosted at the university's network (air-gapped from other resources for safety reasons). It contains several web pages that are vulnerable to SQLi, and some that have the mitigation programmed in. The students utilize this web shop to practice different types of SQLi attacks. For the assignment, the students work in pairs to exploit the SQLi vulnerability to reverse engineer the entity relationship diagram of the back-end SQL database. This links the security lecture with the rest of the course, and makes them realize the extent of information such a simple attack can reveal.

\subsection{Web security in WDT (CSE1500)}
CSE1500 is a Q2 course that is divided into two parts, one focusing on web technologies and the other on databases, taught by two different instructors. 
Over five weeks, the students learn about web programming fundamentals in the web technologies part of the course. 
We cover the \texttt{Connection Security}, and \texttt{Software Security} KAs, and include a lecture on \textit{Web Security} in week 4. 
The LO is to understand the prevalence of security vulnerabilities in web applications, especially the OWASP Top-10. In a 90-minute lecture, the students learn to exploit each vulnerability on an publicly available instance of the OWASP Juice Shop \cite{owasp2023} -- a vulnerable web application provided by OWASP for educational purposes. They also learn to patch each of these vulnerabilities.  

\section{Evaluation and Analysis}

\begin{table*}[t]
\caption{Post-course survey responses from first, second, and third year BSc students.}
\label{tab:survey-responses}
\resizebox{2\columnwidth}{!}{%
\begin{tabular}{clccccc}
\toprule
\textbf{No.} & \multicolumn{1}{c}{\textbf{Statement}} & \textbf{\begin{tabular}[c]{@{}c@{}}Strongly \\ Agree\end{tabular}} & \textbf{\begin{tabular}[c]{@{}c@{}}Somewhat \\ Agree\end{tabular}} & \textbf{\begin{tabular}[c]{@{}c@{}}Neither Agree \\ Nor Disagree\end{tabular}} & \textbf{\begin{tabular}[c]{@{}c@{}}Somewhat \\ Disagree\end{tabular}} & \textbf{\begin{tabular}[c]{@{}c@{}}Strongly \\ Disagree\end{tabular}} \\ \toprule
S1& The security lecture(s) were interesting & 66 & 24 & 5 & 2 & 0 \\ 
S2& The security lecture(s) were relevant to me & 42 & 42 & 9 & 2 & 2 \\ 
S3& I understand the relevance of security for the course & 82 & 8 & 4 & 1 & 2 \\ 
S4& [...] increased my interest in learning about other security topics & 50 & 24 & 15 & 8 & 0 \\ 
S5& Live demo improved my understanding of attacks and their mitigation & 56 & 23 & 14 & 3 & 1 \\ 
S6& The security content was aligned with the course & 40 & 39 & 10 & 7 & 1 \\ 
S7& Combining security with CS helped me consider security-by-design & 45 & 39 & 9 & 3 & 1 \\ 
S8& [...] helped me think like a security practitioner & 29 & 39 & 20 & 7 & 2 \\ 
S9& [...] helped me think like a cyber attacker & 40 & 36 & 14 & 5 & 2 \\ 
S10& I would prefer relevant security lectures in CS courses & 53 & 22 & 12 & 6 & 4 \\ 
S11& I would prefer a dedicated security course & 54 & 23 & 12 & 6 & 2 \\ \bottomrule
\end{tabular}}
\end{table*}

It is challenging to evaluate a crosscutting intervention since the security content is divided across various courses.
We obtained quantitative and qualitative feedback from the three stakeholders involved in our intervention -- we conducted a post-course survey with students who followed the security lectures, and semi-supervised interviews with course instructors. We also discuss the experience of the security expert in the context of the interview questions.
For this, we obtained the necessary approvals from our institutional IRB.

\paragraph{Student survey.} The survey started with an information sheet where the participants were informed about the purpose of the study and its voluntary and anonymized nature. We did not collect any personally identifiable information regarding the participants. 

We adapted the survey from Grosz \etal \cite{grosz2019embedded}. Table \ref{tab:survey-responses} shows the 11 statements presented to the students, which they were asked to rate on a 5-point Likert scale. The students were requested to fill out the survey right after the \textit{Data Security} lecture in the last run of IDM. We also contacted year-2/3 students to fill out the same survey since they had undergone the complete intervention, and would have a better perspective on it.

For the students who liked the crosscutting intervention, we further asked the reason behind their preference in a free-response textbox. For the students who preferred a dedicated security course, we asked what they expected from such a course. We also asked their least and most favorite aspects of the security lecture. Finally, we asked in which other year-1 courses they would like to see more security content. We analyzed these responses using an inductive thematic coding process \cite{braun2006using}.

\paragraph{Instructor interview.} The course instructors were invited via email to participate in a semi-supervised interview. The interviews were conducted by the author remotely via Zoom. Each interview lasted approximately 20-25 minutes. The interview was recorded, following the consent of the course instructor. 
The instructor was asked about the process of including cybersecurity in their course, their thoughts on the security content in their course, whether the security content competed with other topics `more' related to their course, and their recommendation on successfully incorporating cybersecurity as a crosscutting concept across a CS curriculum. These responses were transcribed and analyzed using an inductive thematic coding process.

\subsection{Student perception}

We received 97 responses from the students: 80 from year-1, 13 from year-2, and 4 from year-3 students. 
Over 85\% of the participating students found the security lecture engaging (S1), relevant (S2), and understood its purpose in the respective course (S3). They particularly liked the real-world examples in the lecture. 74 (76.3\%) students were interested in learning about additional security topics (S4). 
70 out of 80 (87.5\%) year-1 students thought that the security lecture was well-embedded within the IDM course (S6).
A year-1 student shared: \textit{``this [SQLi] lecture was easy to grasp for [me] now because we have been dealing with SQL databases in this course''.}

72 out of 80 (90\%) year-1 students found that the practical hands-on component of the lecture helped them understand the SQLi attack and the use of prepared statements as a countermeasure (S5). 
In fact, the live demo was the most frequently mentioned (and impactful) aspect of the lecture, \eg a year-1 student responded \textit{``[t]hat we could also do the SQL injections ourselves, so [the lecture] was more interactive''}, while another student responded \textit{``the demo in which [we] `hacked' (SQL injected) a server. Honestly it was very good, I have nothing to add to it, other than that I would want more''.}
What we found particularly interesting was that even the year-3 students (who had not followed a security course for two years) remembered the live demo as being the aspect of the course that made them recognize the importance of cybersecurity, \ie \textit{``the demo [made the attacks] suddenly [feel] real and too easy''. }

While 84 (86.6\%) students agreed that combining CS topics and their security implications within a single course helped them consider security-by-design (S7), there was a negligible difference between student preferences for a crosscutting intervention \textit{vs.} a dedicated security course. 77.3\% students agreed with S10 and 79.4\% agreed with S11. This suggests that the students were generally more enthusiastic about security education. 65 students preferred both settings; 3 students preferred security as a crosscutting concept; 6 students preferred a dedicated security course; and 2 students did not share their preference. 

The students believed that it is more impactful to include relevant security topics when the CS topics are fresh in their minds, compared to having a security course far away in the curriculum. One year-3 student shared that they would like security as a cross cutting concept rather than a dedicated security course, since \textit{``[the] knowledge of a certain course can be directly applied''.} A year-1 student suggested that \textit{``[this way, we] can cover more of the different material related to security and link it to different topics''.} Similarly, a year-2 student added that \textit{``it would help connecting [security] to other topics in the curriculum more easily, while [we] are already working with those topics''.} 
In contrast, the students who preferred a dedicated security course wanted discussions on additional security topics in more depth, \textit{``a mix of practice and theory''}, and practical CTF-style assignments. These options do not necessarily have to be mutually exclusive in our opinion, as discussed in Section \ref{sec:rec}.

One of our goals was to inculcate the security mindset by including both offensive and defensive security components in the lectures. 65 students agreed that the lecture helped them think like both a security practitioner (S8), and a cyber attacker (S9). The students shared that \textit{``learning [how] to hack is important because you will learn how [attackers] think if you tried it yourself''}, and 
\textit{``Doing is the best way to learn how to protect [systems]''}. 

Finally, the top-4 courses where the students wanted to follow security lectures were (in order of preference): Computer Networks (CN), Software Quality and Testing (SQT), Object Oriented Programming (OOP), and Computer Organization (CO). Since the survey was conducted in Q3, the majority of the (year-1) students were unaware that SQT was already part of our intervention, and CN included some security elements (both are Q4 courses). 

\subsection{Instructor reflection}\label{sec:instructor-reflection}
While we invited all the instructors involved in our intervention, we were only able to interview the instructor of Software Quality and Testing (SQT) -- 
Dr. Maurício Aniche (MA).

The security content in SQT reduced over the years (from a 180-minute lecture in 2018 to a 45-minute lecture in 2023). This was because the course contents evolved, \eg due to the introduction of MA's textbook \cite{aniche2022effective}, which did not include security.

MA shared his skepticism on successfully maintaining security as a crosscutting concept, suggesting that \textit{``[it] is beautiful on paper but does not work well in practice''}. 
He postulated that the security lecture would have been better connected to the course if the responsible instructors had been teaching it rather than outsourcing it to the security expert -- \textit{``We could [teach] it easily. It would be a different lecture, for sure. We would focus on security issues that we face as normal software engineers''.} 
Moreover, there were logistical challenges with outsourcing lecture material, \eg the security expert was not connected to the day-to-day of the course, so they were unlikely to answer student questions about the topics discussed in prior lectures. Evaluation was also difficult since the security expert had insufficient budget to create security questions that integrated with the automated test generation toolkit used in the exam.  

MA suggested that this set up may potentially work with a more interactive relationship between the course instructor and the security expert, \ie the course instructor takes an active role in suggesting relevant security content, such that it fits with the overall story line of the course. 
In practice, maintaining such an interactive relationship is difficult since the responsible instructor is more focused on topics that are directly related to the course -- \textit{``We have 9 weeks, and we have to pick what to teach. There are more important things to talk about than security. In a basic testing course, should we talk about mocks or security? I would say mocks because developers will write more mocks than security tests''.}

He advocated for a dedicated cybersecurity course so the students could learn it properly, as opposed to learning about it in different courses where it is never a priority. 
This also speaks to the misaligned priorities from an organizational standpoint that have not yet established a compulsory security course. 

Nevertheless, MA described the crosscutting intervention in SQT as \textit{``a positive experience''}, sharing that the security lecture evolved in the right direction over the years, it was well-contextualized with concepts familiar to the students, and the Pacman assignment was fun for the students. 
In order to enhance the connectedness of the security lecture with the course content, MA suggested that either the security expert attends the other lectures and highlights when a topic is related to cybersecurity, or the responsible instructor also learns about the security topic so they can highlight the link themselves. Nevertheless, both stakeholders must meet in the middle for a seamless infusion of security in the course.

\subsection{Security expert experience}

The initial implementation of the learning materials took about a year. It is difficult to design a vulnerable web application that the students can compromise in exactly the ways the instructor expects them to. Regardless, the learning materials are an investment -- they are expensive to design at first, but can be reused with minimal effort for several years. 
It is also noteworthy that the materials cannot remain static for long -- they must be updated more frequently than other course materials as the threat landscape evolves. 

A significant amount of time was spent negotiating with the course instructors to allocate sufficient space for the security content. As mentioned by MA, part of it had to do with topic prioritization. The other part was giving insufficient importance to cybersecurity, \eg the instructor of IDM disagreed with the syntax of an SQL injection command because he believed it was an invalid way to write an SQL query, even though it caused a successful attack. This highlights the intrinsic difference in the mentality of security \textit{vs.} non-security instructors.
Finally, it was complicated to evaluate the crosscutting intervention since the feedback obtained through pre-existing course evaluation forms was disconnected. 

These bottlenecks made the crosscutting intervention a significantly more expensive task than managing a security course alone. However, the students loved the security content sprinkled across many courses. 
In this way, the role of the security expert almost felt like that of an advocate, negotiating on behalf of the students.

The lack of organizational oversight, \eg by a security coordinator to ensure topic linkage, made it difficult to sustain the crosscutting intervention across different courses over several years. This is evident from the diminishing amount of security content over time. 
However, this was not the case for all security content: The \textit{Web Security} lecture in WDT was developed during the design phase of the course, which made it fit seamlessly with the rest of the course. Similarly, the limited security content in CN was developed by the responsible instructor, who also happened to have a security background. These examples suggest that there \textit{are ways to successfully incorporate} security in a course. Ultimately, instructor willingness and an active communication between them and the security expert were the biggest factors in sustaining the security education. 


%

\section{Recommendations}\label{sec:rec}

Based on our experience over a five-year period, student perceptions, and instructor reflections, we recommend the following modifications to our intervention:
\textit{1) Organizational oversight}: A crosscutting intervention across an entire curriculum is definitely not a one-person job. While we show that such an intervention is possible and is appreciated by the students, it requires a lot of maintenance that cannot be done without organizational support. We recommend involving a `security coordinator' and multiple security experts. The security coordinator collaborates with the security experts and course instructors for topic selection. The security expert remains responsible for teaching it, and the instructor for ensuring its proper integration in the course. The security coordinator ensures that the different security lectures have a flow, and revisit important concepts regularly, \eg as a learning path.
\textit{2) Intervene during course (re-)design}: The course instructor and the security expert must work in unison to select apt security content and familiar terminology, and must maintain frequent communication to monitor the progress of the intervention. Feedback must be collected from the three stakeholders and incorporated in subsequent iterations of the course or during course re-design. 
\textit{3) Hybrid intervention}: It may be useful to have a combination of a compulsory crosscutting intervention (for a basic introduction and interest generation), and a dedicated security course in year-2 or year-3 (for in-depth topical coverage).
If curriculum allows, it may even be possible to have a dedicated (compulsory) introductory course in year-1 that teaches basic security concepts in an abstract way. In later courses, those concepts can be revisited (in a crosscutting intervention) with more concrete knowledge of computing topics.

\section{Final remarks}
This paper describes our practical experience of integrating cybersecurity as a crosscutting concept 
in an undergraduate CS program, which is useful in case of limited space in the curriculum.
The security topics are selected based on underrepresented knowledge areas in CSEC2017, and are taught by a security expert.
The security expert fine-tunes the learning materials together with the responsible course instructor to seamlessly fit the story line of the course. 
The key characteristics of our intervention are: offensive and defensive security learning objectives, hands-on practical exercises, and examples of real-world exploits. 

We studied the complex dynamic between the three stakeholders: the course instructor, the security expert, and the students. 
We found that seamlessly infusing and sustaining security within the story line of existing courses is possible but requires organizational support and the cooperation of several stakeholders.
The discrepancy between the stakeholder incentives was the main force of resistance in this intervention: While the students loved security lectures, the course instructors would rather teach their own content, either due to topical trade-offs or insufficient exposure to security threats. Thus, the security expert had to make a strong case for the impact of integrating cybersecurity in their courses.  

Based on our five-year experience, we believe there is merit in integrating cybersecurity as a crosscutting concept within existing courses, and would strongly encourage educators to support these efforts to meet the needs of a cyber-ready workforce. 

\subsection*{Acknowledgments} We thank Prof.dr. AE Zaidman for his guidance. We also thank Junaid Mehmood and the anonymous reviewers for their feedback.

\bibliographystyle{ACM-Reference-Format}
\balance
\bibliography{_teach}
  
\end{document}